# Towards Causality Extraction from Requirements

Jannik Fischbach, Benedikt Hauptmann, Lukas Konwitschny, Dominik Spies
Qualicen GmbH, Germany
{firstname.lastname}@qualicen.de

Andreas Vogelsang
Technische Universität Berlin
andreas.vogelsang@tu-berlin.de

***Abstract*—System behavior is often based on causal relations between certain events (e.g. If $event_1$, then $event_2$). Consequently, those causal relations are also textually embedded in requirements. We want to extract this causal knowledge and utilize it to derive test cases automatically and to reason about dependencies between requirements. Existing NLP approaches fail to extract causality from natural language (NL) with reasonable performance. In this paper, we describe first steps towards building a new approach for causality extraction and contribute: (1) an NLP architecture based on Tree Recursive Neural Networks (TRNN) that we will train to identify causal relations in NL requirements and (2) an annotation scheme and a dataset that is suitable for training TRNNs. Our dataset contains 212,186 sentences from 463 publicly available requirement documents and is a first step towards a gold standard corpus for causality extraction. We encourage fellow researchers to contribute to our dataset and help us in finalizing the causality annotation process. Additionally, the dataset can also be annotated further to serve as a benchmark for other RE-relevant NLP tasks such as requirements classification.**

## I. Introduction

Functional requirements usually specify a system from three perspectives: (1) the inputs to be processed by the system, (2) the expected system behavior once these inputs occur, and (3) the outputs that the system shall produce. Consequently, they often follow a causal pattern (e.g. If $e_1$, then $e_2$). Extracting causal relations from texts enables several analytical disciplines and is already used for event prediction [1] and information retrieval [2]. However, the potential of extracting causal knowledge embedded in requirements has not yet been leveraged for Requirements Engineering (RE). In this paper, we shed light on the added value that causality extraction can provide to RE by presenting two use cases: derivation of test cases and dependency detection between requirements. Since existing methods [3] fail to extract causal relations with reasonable performance, causality extraction remains an open NLP problem. In this paper, we describe first steps towards building a new approach for causality extraction. Specifically, we propose an NLP architecture based on Tree Recursive Neural Networks (TRNN) that we will train to identify causal relations in NL requirements. As the RE community currently lacks a suitable training corpus for neural networks, we initiated the creation of a gold standard corpus and extracted 212,186 sentences from 463 publicly available requirement documents. In this paper, we present the dataset[1] and a causal annotation scheme for labeling the dataset to use it for training and evaluating the TRNN.

[1] Available at https://figshare.com/s/725309c06b9dc82aa4a1. Due to the terms of use of some sources, we can only share the URLs of the collected documents. We attached a script to download the dataset automatically.

## II. Terminology

We first need to define the concept of causality and clarify what is meant by causality in requirements. Causality has been studied in many different disciplines as Psychology [4] and Philosophy [5]. All definitions of causality have in common that a causal relation denotes the relation between two events $e_1$ and $e_2$, where $e_1$ causes the occurrence of $e_2$. In this context, $e_1$ is called cause and $e_2$ effect. Causality can occur in two different degrees and is therefore formally defined as either an implication $\implies$ or an equivalence $\iff$. This can be illustrated by the following example:

REQ 1: '*If the system detects an error, an error message shall be shown.*'

REQ 1 consists of the cause "the system detects an error" ($c_1$), and the effect "an error message shall be shown" ($e_1$). Literally, this statement can be interpreted logically as an implication ($c_1 \implies e_1$), in which $c_1$ is a *sufficient* condition for $e_1$. Interpreting REQ 1 as an implication, requires the system to display an error message if $c_1$ is true. However, it is not specified what the system should do if $c_1$ is false. The implication allows both the occurrence of $e_1$ and its absence if $c_1$ is false. The case of $c_1 = \mathit{false}$ is *underspecified*. From a testing point of view, underspecified requirements can be problematic, because the negative scenario is not specified. In fact, when reading REQ 1, it may as well be reasonable to assume that the error message shall not be shown if the error has not been detected. This interpretation corresponds to a logical equivalence ($c_1 \iff e_1$), where $c_1$ is both a *sufficient* and *necessary* condition for $e_1$. Even if not literally correct, a number of approaches interpret the literal of REQ 1 as an equivalence. For example, Mavin et al. [6] proposes to treat a causal relation in requirements as equivalence in order to avoid ambiguity. Specifically, they define causality in two of their EARS patterns: Unwanted Behaviour and Event Driven. In both patterns, "the system is required to achieve the stated system response if and only if the preconditions and trigger are true" [6]. This makes the requirement clear and testable, because both positive and negative scenarios are defined. Interpreting REQ 1 as an equivalence, requires the system to display an error message if and only if it detects an error. This reflects, as we argue, the system behavior we implicitly want to express with the requirement even if we omit the "only if" phrase. In practice, there are often several requirements describing the same effect:

REQ 2: '*If the user enters a wrong password, an error message shall be shown.*'

REQ 2 specifies a further scenario (i.e. a wrong password ($c_2$)), in which the system should issue an error message. In such cases, both requirements must be formalized by a joint equivalence relation, i.e. $c_1 \vee c_2 \iff e_1$. In the remainder of this paper, we refer to an equivalence relation between $c_m$ and $e_n$ when we mention causality in requirements. If several requirements define the same effect, we assume that each requirement describes a separate case in which the effect should occur. We therefore link their causes with disjunctions and build a single equivalence relation.

## III. Use Cases in Requirements Engineering

We describe two use cases in which automatic causality extraction from requirements supports RE activities and discuss the necessary performance level of an automatic approach.

*1) Automatic Test Case Derivation:* Deriving relevant test cases from requirements is a laborious activity, which accounts for 40–70% of the total effort in testing [7]. Currently, test cases are mostly derived manually due to a lack of tool support [8]. Existing approaches for automatic test derivation require semi-formal [9] or even formal notations [10] of requirements. However, unrestricted natural language (NL) is prevalent in practice [11]. To derive test cases from NL requirements, the specified system behavior and the combinatorics behind the requirement need to be understood. Specifically, we need to understand the embedded causal relation to determine the correct combination of test cases that cover all positive and negative scenarios. This can be illustrated by the following requirement: *"If the customer is older than 23 years and shows a valid driving license, the system does not charge an increased fee."* The requirement contains two causes '*customer is older than 23 years*' ($c_1$) and '*[customer] shows a valid driving license*' ($c_2$) and one effect '*the system does not charge an increased fee*' ($e_1$). The causes are connected by a conjunction indicating that the effect depends on the occurrence of both causes, i.e. $c_1 \wedge c_2 \iff e_1$. Currently, practitioners have to extract the causal relation manually and determine the combinations of causes and effects that need to be covered by test cases. This is not only cumbersome but also becomes increasingly error-prone with growing requirements complexity as the number of potential test cases increases by $2^n$, where $n$ is the number of causes. Therefore, test cases may be missed during manual creation or testing effort may be spent on irrelevant test cases.

We argue that causality extraction combined with existing automatic test case derivation contributes to the alignment of RE and testing. For example, Fischbach et al. [12] show how extracted causal relations are mapped to a Cause-Effect-Graph from which test cases can be derived automatically. This leads to time savings as the expected system behavior is interpreted automatically, and to a better test coverage as the required test cases are determined by heuristics. Hence, we argue that causality extraction represents one of the last remaining puzzle pieces on the way to a fully automated test case generation from NL requirements. For this use case, we need causality extraction with high precision because false positives may result in invalid test cases. A low(er) recall, on the other hand, results in missing test cases, which is also not good but is closer to the current reality. In any case, a careful assessment of precision and recall needs to be made also considering the human's ability to identify false positives (cf. [13], [14])

*2) Automatic Dependency Detection between Requirements:* With the increasing complexity of modern systems, both the complexity of requirements and their quantity increase. Additionally, as requirements are primarily expressed by unrestricted natural language, it is difficult to keep an overview of all requirements and their relations [15]. Undetected redundancies and inconsistencies within the requirements may lead to faults in the system design [16]. Furthermore, knowledge about requirements relations is essential to understand the impact of a proposed software change on the overall system functionality [17]. To better understand the relations between requirements, different dependency models have been proposed describing relations on different abstraction levels. An integrated view is provided by Dahlstedt and Persson [18].

We argue that an automatic causality extraction from requirements can help to compare the semantics by analyzing the different embedded causal relations. As a result, relations between requirements can be identified automatically. Causality extraction contributes to the detection of 4 out of the 7 dependency types described by Dahlstedt and Persson [18]:

*a) Contradictory Requirements:* The causal relations *contradict* each other, i.e. the content of two requirements conflict with each other. Such conflicts may lead to inconsistencies in the system design. Identifying these enables engineers to identify requirements that need to be re-discussed with the stakeholders to clarify the desired behavior.

Example: $c_1 \iff e_1$ contradicts $c_1 \iff \neg e_1$.

*b) Requirements That Require Others:* If the effect of a causal relation $r_1$ appears as cause in another causal relation $r_2$, the fulfillment of $r_2$ depends on $r_1$. In other words, the requirement that includes $r_2$ *requires* the requirement containing $r_1$. Indicating this kind of relation helps engineers to maintain requirements traceability. If $r_1$ is changed, the engineer is aware of which other requirements are affected.

Example: $e_1 \iff e_2$ requires $c_1 \iff e_1$.

*c) Redundant Requirements:* The causal relations are congruent, i.e. the two requirements describe the same system behavior and are therefore *redundant*. Engineers may remove one of the requirements to keep the requirements suite minimal.

Example: $c_1 \iff e_1$ is congruent to $c_1 \iff e_1$.

*d) Requirements Refinement:* A causal relation $r_1$ that specifies a subset of the solution of another causal relation $r_2$ indicates that $r_1$ *refines* $r_2$, i.e. the system behavior described in one requirement is defined more precisely in another.

Example: $c_1 \iff e_1$ refines $c_1 \wedge c_2 \iff e_1$.

## IV. Problem Statement & State of the Art

Causality occurs in different forms. Blanco et al. [24] distinguishes between: a) Marked and unmarked causality, b) ambiguous and unambiguous cue phrases and c) implicit and explicit causality. A causal relation is *marked* when a specific phrase indicates causality, e.g. "The system restarts because of an error." is *marked* by the cue phrase "because", while "Drive slowly. The highway is iced." is *unmarked*. However, it is not always feasible to infer causality from a certain cue phrase. Some phrases, e.g. "since", may indicate causality, but they also occur in other contexts (e.g. to specify time constraints). These phrases are denoted as *ambiguous* while phrases that always indicate causality (e.g. "because"), are defined as *unambiguous*. Causal relations that contain information about both causes and effects are called *explicit* (e.g. "The pump starts when the water level rises."). In *implicit* relations, the causes and effects are not explicitly mentioned. "A parent process kills a child process." is *implicit* since the effect, "the child process is terminated", is not explicitly stated. These examples indicate the challenges for automatic causality extraction.

Tab. I: Existing Techniques for Causality Extraction from Natural Language.

| **Extraction Level** | **State-of-the-Art** | **Example:** *If A is true and B is false, the system shall show an error message.* | **Relation** |
|---|---|---|---|
| Word | Chang and Choi [19], Rink et al. [20] | true $\iff$ message<br>false $\iff$ message | $r_1$<br>$r_2$ |
| Phrases | Dasgupta et al. [21], Li et al. [22] | A is true $\iff$ system shall show an error message<br>B is false $\iff$ system shall show an error message | $r_3$<br>$r_4$ |
| Full (Our Scope) | TRNN - Socher et al. [23] | A is true AND B is false $\iff$ system shall show an error message | $r_5$ |

Several approaches for causality extraction have been developed [3]. However, existing methods fail to extract causality with reasonable performance. Rule-based approaches [25], [26] extract causality by applying linguistic patterns such as "[cause] and because of this, [effect]". Since these approaches search for specific words, they are limited to the detection of *marked* causality. In addition, their performance relies on the hand-coded patterns, which requires extensive manual work. Other approaches use Machine Learning (ML). Chang and Choi [19] use a Naive Bayes classifier to predict the probability of a causal relation given a certain cue phrase (e.g. causative verb). Rink et al. [20] propose a Support Vector Machines classifier trained on contextual features (e.g. dependency parse). However, these approaches extract causality only at word level (see extracted relations $r_1$ and $r_2$ in Tab. I). Hence, valuable information about the causes and effects is lost making these approaches unsuitable for our use cases.

Recent approaches address this problem and identify causality on phrase level (see Tab. I). Dasgupta et al. [21] and Li et al. [22] use bidirectional Long Short Term Memory models that outperform existing baseline systems. However, their performance depends on the used training corpora, so that the application of neural networks is usually domain-specific. They use e.g. the BBC news article set [27] not originating from software engineering. We hypothesize that the approaches are therefore difficult to transfer to RE. This matches the finding of Ferrari et al. [28] that requirement documents often exhibit a specific vocabulary, requiring NLP approaches to be trained on RE data to make them usable for requirements analysis. In addition, both approaches only extract cause-effect pairs, whereby the combinatorics between the causes and effects gets lost during the extraction (see $r_3$ and $r_4$). However, we need to extract the entire embedded causal relation to make it useful for test case derivation and dependency detection between requirements (see $r_5$). Thus, we require a new approach to implement our described use cases.

## V. Our Approach

We want to build a new approach for causality extraction based on Tree Recursive Neural Networks (TRNN) [23]. We focus on Deep Learning (DL) techniques instead of shallow ML approaches for two reasons: First, shallow ML approaches rely on carefully extracted features, which requires considerable human effort and time. Manually selecting textual features is challenging and error-prone, especially for a complex NLP task like causality extraction. In contrast, DL techniques identify features automatically based on their capability of representation learning. Second, multiple studies [29] have shown that DL techniques outperform traditional approaches in NLP tasks that require extensive linguistic skills (e.g. syntactic parsing, machine translation). Since causality extraction requires comprehensive language understanding, we see potential in the application of DL techniques. This section describes why we selected TRNNs from the set of DL techniques and presents how a TRNN can be used to extract causality from requirements by introducing a feasible NLP architecture. DL techniques require a large amount of training data. Since we currently lack a large training corpus in the RE community, we initiated the creation of a gold standard corpus, which we will annotate specifically for causality extraction. This section describes the process of creating this corpus and outlines how the requirements need to be annotated to train the TRNN. We also discuss potential limitations of our approach.

### A. NLP Architecture for Causality Extraction

*a) Principal Idea:* A TRNN is based on the idea that natural language can be understood as a recursive structure. For example, the syntax of a sentence is recursively structured, with noun phrases containing relative phases, which in turn contain further noun phrases, and so on. Recovering this recursive structure enables to better understand the composition of a sentence. Socher et al. could already prove that a TRNN is suitable for syntactic parsing of NL sentences [23]. We transfer this idea to causality extraction and understand a causality relation as a recursive structure. We argue that a causal relation consists of causes and effects, which in case of conjunctions and disjunctions consists of further causes and effects, and so on. This results in a tree-like structure of cause and effect nodes forming the full requirement (see Fig. 1). By recovering this tree-like structure, we do not lose the combinatorics between the causes and effects which allows us to entirely extract the causal relation and use it for the described use cases.

*b) Input Representation of Requirements:* To process requirements with a neural network, their words must be converted into a word embedding. In recent years, several methods have been developed to implement word embeddings. Traditional methods like word2vec [30] are capable of transforming a word into a single vector representation, however, they do not consider the context of the word. Hence, words are always represented as the same vector, although they can have different meanings depending on their context. To address this problem, contextual word embeddings like ELMO [31] and BERT [32] were developed. Since BERT shows clearly better performance than ELMO in previous studies [33], we will use BERT embeddings in our proposed architecture.

*c) Causality Extraction:* We define a requirement as a sequence of vectors $(v_i, ..., v_n)$ and aim to determine which vectors describe a cause or effect and to exclude non-causal vectors. Hence, we specify a cause and effect as a segment of a requirement. A TRNN can be trained to identify such segments from NL sentences. For this purpose, the TRNN builds a binary tree, i.e. it always considers two adjacent vectors and learns to merge them if they belong to the same segment. As a result, the individual segments are extended by one vector in each recursion. At the beginning of each recursion, the TRNN determines a set of all adjacent vector pairs $C = \{[v_i, v_j] | v_i \text{ and } v_j \text{ are adjacent}\}$. In order to determine which pair should be merged, the neural network calculates a score $s_{i,j}$ for each adjacent vector pair $[v_i, v_j]$. After computing all scores, the TRNN selects the pair with the highest score and merges both vectors to one parent vector $p$. Each $p$ is assigned a label $l$, which is intended to define the segment. In Fig. 1, for example, the adjacent vector pair $[v_2, v_3]$ is merged to a parent vector with the label *VAR*. The training of the network aims at increasing the scores for content related pairs, which should be combined into a segment, and at decreasing the scores for non related pairs. After each recursion, the set of possible pairs $C$ is updated and the scores of the new potential pairs are calculated. This process is repeated until all pairs are merged. For detailed calculation of the scores, please refer to Socher et al. [23]. Fig. 1 illustrates this process for an exemplary requirement and the corresponding binary tree, which we will implement using a TRNN. To extract the causes and effects precisely and to reflect their combinatorics, we will train the TRNN to identify the following segments:

- **S**: This label is assigned to the root node of the tree and defines the full sentence resp. requirement.
- **CR**: This label is assigned to the segment that contains all causes and effect segments. Consequently, it contains all vectors that are part of the causal relation.
- **C**: This label is assigned to a segment, which contains a single cause or further cause segments.
- **E**: This label is assigned to a segment, which contains a single effect or further effect segments.
- **CON**: This label is important to consider two causes or effects connected by a conjunction during the extraction.
- **DIS**: This label describes a disjunctive relation between two causes or effects.
- **V**: Causes and effects can be split into further segments. For example, the cause "variable A is true" shown in Fig. 1 can be divided into a variable "variable A" and condition "is true". This splitting is useful as it enables us to get a better understanding of the content of the causes and effects. In addition, the extracted causes and effects can be used ideally for the described use cases. For example, inputs and outputs of a test case are always in the form variable and condition.
- **CD**: This label indicates the condition in a cause or effect.
- **P**: Often two vectors are not sufficient to describe an entire segment. In Fig. 1, for example, the condition of the effect consists of 5 vectors. For such cases, we introduce the label P as "part of" and use it to denote segments that belong to a larger segment.
- **c**: This label is applied only on word level and indicates vectors that are part of the causal relation.
- **nc**: This label is also only applied on word level and indicates vectors that do not contain any content about the causal relation.

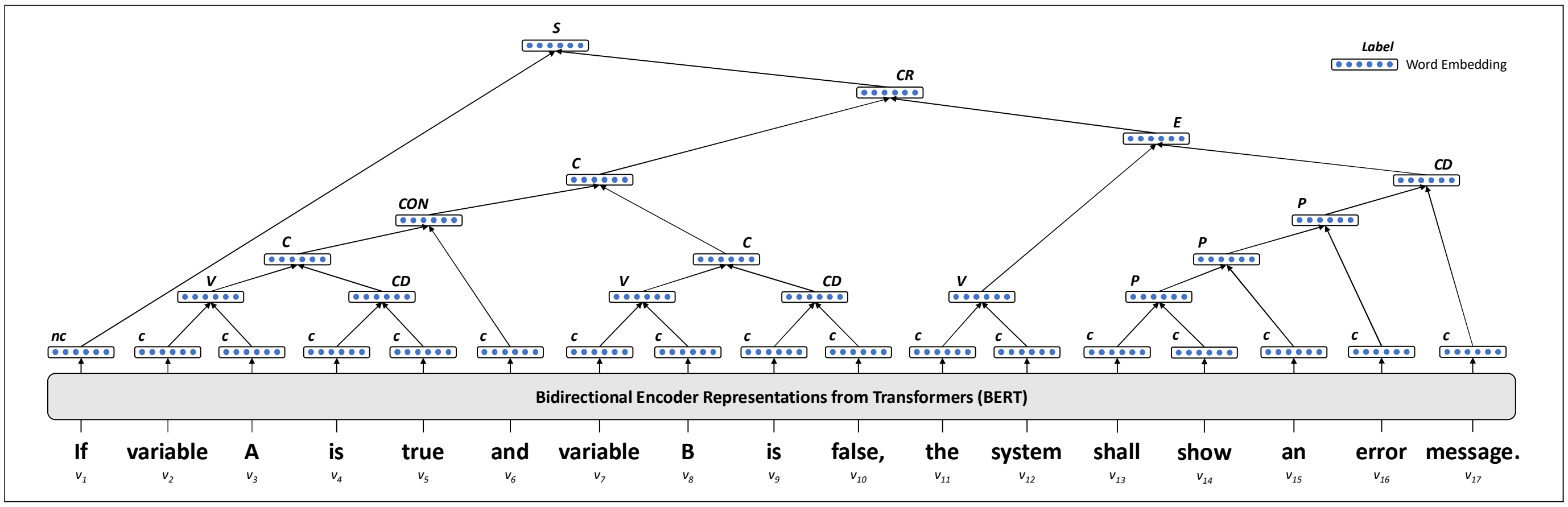

Fig. 1: Our proposed NLP architecture for causality extraction from requirements. First, each word of the requirement is converted into a BERT embedding. Second, the TRNN merges related adjacent vector pairs to parent vectors. By this, the individual segments (e.g. causes) of the sentence become visible resulting in a binary tree. The structure of the embedded causal relation is depicted by the segment CR.

### B. *Resource Creation*

In order to train the presented architecture, we need a large labeled training corpus. To the best of our knowledge, there is no corresponding corpus available in the RE community. The PURE dataset [28] is a first contribution in this direction, however, it is not suitable for our purposes: First, it does not contain enough sentences to train neural networks but is rather suitable for training traditional ML approaches. Secondly, it is not labelled yet and thus can not be used for supervised causality extraction. To pave the way for DL approaches into the RE community, we initiated the creation of a large gold standard corpus of requirements. In the following, we describe the process of creating the corpus and explain how we will annotate the requirements to train our proposed architecture. The corpus is not only intended to be used for causality extraction, but should also serve as the basis for further studies in the RE community. We welcome fellow researchers to use the dataset as a benchmark for other RE relevant NLP tasks.

*a) Data Collection:* We collected publicly available requirements specifications by means of a web search. We queried Google and libraries as Everyspec to retrieve documents from different domains. Tab. II summarizes the searched websites and the applied search terms. We only considered documents which are in PDF format, have at least 10 pages, are written in English, and do contain requirements. To verify the latter, we conducted a brief manual review of each document. Our search led to the identification of 463 requirement documents. The shortest document has 10 pages, while the longest contains 2822 pages. On average, a document has 88 pages.

Tab. II: Overview of collected requirement documents.

| Website | Search Terms | # Docs |
|---|---|---|
| google.com | requirements document (pdf) OR requirements specification (pdf) | 124 |
| ntrs.nasa.gov | requirements document | 11 |
| sci.esa.int | requirements document | 7 |
| www.eurocontrol.int | requirements | 50 |
| www.everyspec.com | requirements OR system requirements | 178 |
| www.etsi.org/standards | most recent | 93 |

*b) Preprocessing:* In order to make the data included in the documents usable for training, we have to extract complete sentences. Simply extracting the lines and training on incomplete sentences is not reasonable as essential phrases of the sentences are neglected. Additionally, we need to clean up the data, i.e. we need to remove lines that are only used for structure purposes (e.g. headings) and do not contain RE relevant content. To this end, we performed the following steps:

1) Extract raw text lines from the PDF file.
2) Preprocess the resulting lines by removing leading and trailing white spaces and compressing a group of consecutive white spaces to one.
3) Filter out lines that contain clearly no content. A line is filtered if at least one of the following criteria is satisfied:
    a) The line starts with "Figure" or "Table".
    b) The line starts with a page character or "Chapter".
    c) The line contains less than 50 characters and does not end with ".", "?" or "!".
    d) The line contains a group of at least 4 consecutive "." characters (deletion of entries in table of content).
4) Delete enumeration marks like "a)".
5) Build paragraphs by splitting the text at empty lines and combining consecutive lines.
6) Split the paragraphs into sentences.
7) Manually improve the sentence splitting by combining two consecutive sentences if the first sentence ends with "e.g." or "i.e.".
8) Remove the prefix "Note" if it is not followed by "that".

*c) Data Description:* After preprocessing, our dataset contains 212,186 complete sentences. Not every sentence contains a causal relation making them unsuitable for training our approach. Requirement documents contain, for example, many static requirements that describe a state of the desired system and do not cover any input-output combinations. To make the dataset usable for training the TRNN, we need first to identify each sentence that conveys at least one causal relation. Therefore, we define the creation of the training dataset as a two-step labeling problem: First, we need to check each record in the dataset and label it with 0 or 1, where 1 means that the record contains a causal relation and vice versa. Second, we have to mark each causal relation. At present, we are not able to state how many of the extracted sentences contain a causal relation. However, we can already demonstrate that about 40,341 sentences exhibit marked causality. We investigated the occurrence of certain causal cues and found that marked causality is mainly expressed by "if" and "when". This first analysis, though, does not replace the manual investigation of all sentences as it might be subject to false positives and the list of searched causal cues might not be complete. Hence, we still have to examine all sentences manually to obtain a holistic picture of causality in requirements.

*d) Annotation Scheme:* A TRNN requires a strongly structured training input. More specifically, it needs to be trained on binary trees. For this purpose, we need to indicate the specific segments in the requirements, so that the algorithm can learn to identify the structure of a causal relation. Based on the presented segments, we designed an causal annotation scheme. The start and end of a segment is represented by two brackets. Each segment is annotated by a label $l \in \{\mathrm{S, CR, C, E, CON, DIS, V, CD, P, c, nc}\}$ to specify its content. Fig. 2 shows the usage of our annotation scheme.

*e) Annotation Process:* Applying the described annotation scheme is laborious and error-prone, especially for complex requirements since each segment must be marked by brackets. To address this problem, we will use the annotation platform INCEpTION [34], which provides an intuitive annotation user interface. It allows to define individual annotation layers and tag sets. We will create layers for each segment and ask the annotators to mark each segment in the requirement. Finally, INCEpTION merges all layers and produces a unified annotation output. Additionally, it allows to implement a custom exporter. We are currently working on an exporter that automatically generates the brackets for each annotation segment, enabling an efficient and less error-prone annotation process. To ensure the reliability of the annotation, we will assign each annotator both unique and overlapping sentences. Based on the overlapping sentences, we will calculate the inter-annotator agreement using the Fleiss Kappa measure [35].

### C. Potential Limitations

As discussed in Section IV, causality extraction is challenging due to the ambiguity of natural language. The expected performance of our approach in terms of precision and recall is therefore difficult to estimate. However, we can already state that our approach will only be suitable for causality extraction from single sentences and will not work with multiple sentences. In addition, our approach is limited to *explicit* causality and cannot detect causes and effects in *implicit* relations. Since the understanding of *implicit* causality is not always clear, even for humans, we assume that it is also difficult to automate.

## VI. Conclusion & Next Steps

Requirements often describe system behavior by causal relations (e.g. If $A$, then $B$). We want to extract the causal knowledge embedded in requirements and use it to derive test cases automatically and to reason about dependencies between requirements. As existing methods are not capable of extracting causality with reasonable performance, we describe first steps towards building a new approach. Specifically, we propose an NLP architecture based on a TRNN that we will train to extract causal relations from NL requirements. Since the RE community lacks a suitable training corpus, we initiated the creation of a large gold standard corpus of requirements to train neural networks specifically for RE purposes. We extracted 212,186 sentences from 463 publicly available requirement documents and provide, to the best of our knowledge, the currently largest available dataset of NL requirements. We also present an annotation scheme for labeling causal relations in requirements. Currently, we are working on the creation of the training corpus for our proposed architecture. Our first experiments show that the application of the annotation scheme takes 0.5–1 min. depending on the complexity of the sentence. Socher et al. [36] trained a TRNN for sentiment analysis and achieved great performance with a sample of 8,000 sentences. We estimate the effort for creating a similar corpus around 100 hours and plan to involve 4 students. We encourage fellow researchers to help us with the annotation process and use the gold standard corpus for other RE-relevant NLP tasks. In addition, we will also leverage the process of creating the gold standard to quantify the "difficulty" of the task by assessing and comparing the performance of the human raters [13].

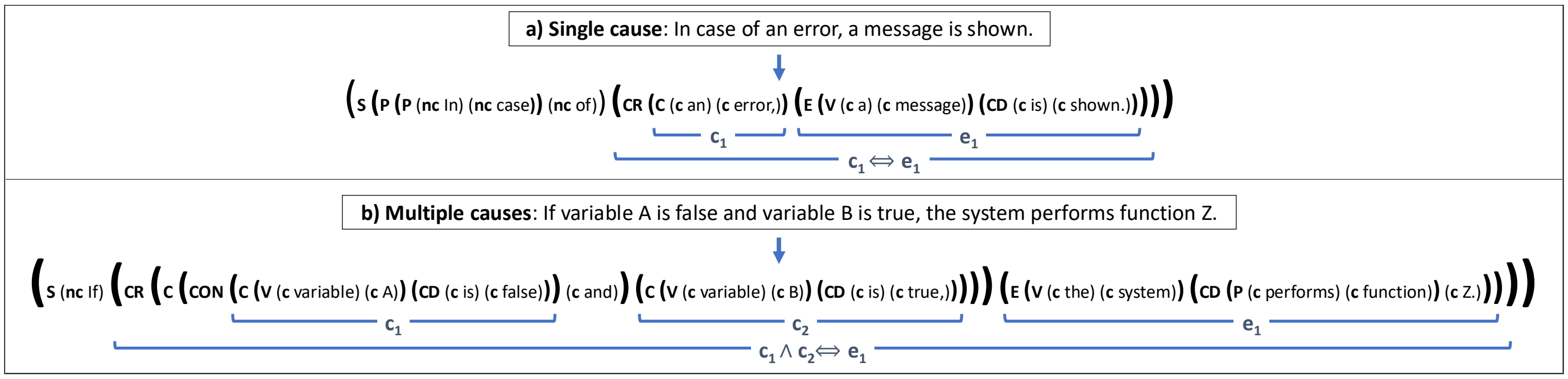

Fig. 2: Annotation examples of two requirements. The different bracket sizes are only used for the sake of clarity.